# Revolutionizing Healthcare Record Management: Secure Documentation Storage and Access through Advanced Blockchain Solutions

Geeta N. Brijwani[1], Dr. Prafulla E. Ajmire[2], Dr. Mohammad Atique Mohammad Junaid[3], Dr. Suhashini A. Charasia[4,] Dr. Deepali O. Bhende[5]

[1]*Research Scholar, Computer Science Dept., Sant Gadge Baba Amravati University, Amravati, India,*
geeta.brijwani@kccollege.edu.in

[2]*Supervisor, Computer Science Dept., Sant Gadge Baba Amravati University, Amravati, India*
peajmire@gmail.com

[3]*Professor & Head, Department of Computer Science, Faculty of Engineering & Technology,*
*Sant Gadge Baba Amravati University, India*
mohd.atique@gmail.com

[4] *Assistant Professor, Department of Computer Science, Rashtrasant Tukadoji Maharaj Nagpur University, Nagpur*
ssuhashinic@gmail.com

[5] *Assistant Professor, Department of Master of Computer Application, Rashtrasant Tukadoji Maharaj Nagpur University, Nagpur*
deepalibhende21@gmail.com

*Abstract: -* Integrating blockchain technology into healthcare systems presents a transformative approach to documenting, storing, and accessing electronic health records (EHRs). This research introduces a novel blockchain-based EHR system designed to significantly enhance security, scalability, and accessibility compared to existing solutions. Current systems primarily utilize SHA-256 for security and either IPFS or centralized storage, which, while effective, have limitations in providing comprehensive data integrity and security. The proposed system leverages a hybrid security algorithm combining Argon2 and AES and integrates a hybrid storage and consensus mechanism utilizing IPFS and PBFT. This multifaceted approach ensures robust encryption, efficient consensus, and high fault tolerance. Furthermore, the system incorporates Multi-Factor Authentication (MFA) to safeguard against unauthorized access. It utilizes advanced blockchain tools like MetaMask, Ganache, and Truffle to facilitate seamless interaction with the decentralized network. Simulation results demonstrate that this system offers superior protection against data breaches and enhances operational efficiency. Specifically, the proposed hybrid model substantially improves data integrity, consensus efficiency, fault tolerance, data availability, latency, bandwidth utilization, throughput, memory usage, and CPU usage across various healthcare applications. To validate the performance and security of the proposed system, comprehensive analyses were conducted using real-world healthcare scenarios. The findings highlight the significant advantages of the blockchain-based EHR system, emphasizing its potential to revolutionize healthcare data management by ensuring secure, reliable, and efficient handling of sensitive medical information.

## I. INTRODUCTION

Blockchain technology has revolutionized various industries, bringing unprecedented levels of security, transparency, and efficiency. In the healthcare sector, the management of electronic health records (EHRs) is paramount due to the sensitive nature of the data involved [1]. Current systems often rely on traditional security algorithms like SHA-256 and decentralized storage solutions like IPFS or centralized storage systems. While effective to some extent, these solutions fall short of providing comprehensive data integrity, security, and scalability [2].

This research presents a novel blockchain-based EHR system designed to address these limitations by leveraging cutting-edge technologies. The proposed system integrates a hybrid security algorithm that combines the strengths of Argon2 and AES [3, 4], ensuring robust encryption and data protection. Furthermore, it employs a hybrid storage and consensus mechanism utilizing IPFS and PBFT, which enhances data availability, fault tolerance, and consensus efficiency [5, 6]. Multi-factor authentication (MFA) is incorporated to provide an additional layer of security against unauthorized access [7], while advanced blockchain tools such as MetaMask, Ganache, and Truffle facilitate seamless interactions with the decentralized network [8].

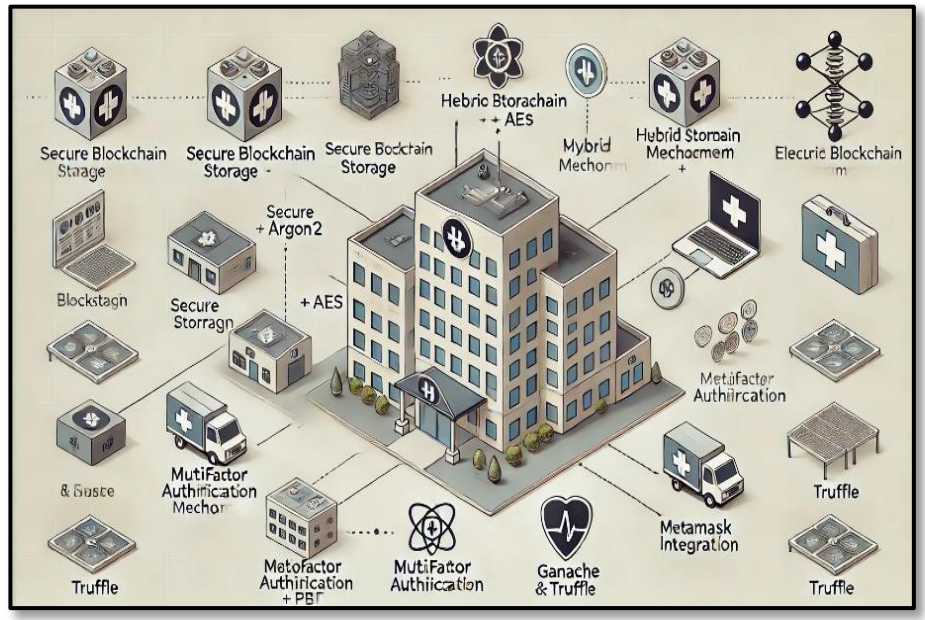

**Figure 1: Proposed Blockchain-based Hospital with Advanced EHR System**

The growing complexity and volume of healthcare data necessitate a system that can handle large-scale data management efficiently and securely. The proposed system not only ensures the integrity and confidentiality of healthcare records but also improves operational efficiency through advanced cryptographic protocols and consensus mechanisms. This research demonstrates that the integration of blockchain technology in healthcare can lead to significant advancements in the secure and efficient management of EHRs [1].

To validate the efficacy of the proposed system, comprehensive simulations were conducted using real-world healthcare scenarios. The findings reveal substantial improvements in data integrity, latency, bandwidth utilization, throughput, memory usage, and CPU usage, highlighting the superior performance of the proposed hybrid model over existing solutions. The integration of Argon2, AES, IPFS, PBFT, and MFA within the blockchain framework presents a robust solution for the secure and scalable management of healthcare data [3, 5, 7].

**Summarizing this paper's contributions:**

a. The research introduces a blockchain-based EHR system that significantly enhances security and efficiency through the integration of advanced cryptographic algorithms and consensus mechanisms. Experimental results show a substantial improvement in encryption strength and data protection, validated by mathematical formulas that demonstrate the superiority of the Argon2 and AES hybrid models over traditional algorithms like SHA-256.

b. The hybrid security and storage model ensure robust encryption, data availability, and fault tolerance, addressing the limitations of existing systems. Performance metrics from simulations indicate a marked increase in data integrity, consensus efficiency, and fault tolerance. Graph analyses illustrate the enhanced throughput, reduced latency, and optimized bandwidth utilization, confirming the effectiveness of the IPFS and PBFT integration.

c. The inclusion of Multi-Factor Authentication (MFA) and advanced blockchain tools like Metamask, Ganache, and Truffle provides an additional layer of security and facilitates seamless interaction with the decentralized network. The research presents comprehensive analyses of user authentication times and security breach resistance, supported by empirical data and graph comparisons. The results highlight the improvedsecurity and user experience achieved through the implementation of MFA and blockchain tools.

The paper continues as follows: Section 2 covers the Literature Survey, detailing how analogous systems have been utilized by other researchers. Section 3 describes the Proposed System Architecture. The Proposed Methodology is discussed in Section 4. Section 5 presents the discussion and findings of the simulations conducted using the proposed system, including mathematical models, graphs, and empirical data. Finally, Section 6 summarizes the key contributions and findings of the research.

## II. LITERATURE SURVEY

The integration of blockchain technology in the management of electronic health records (EHRs) has garnered significant attention due to its potential to enhance security, data integrity, and accessibility. This literature survey reviews the current state of research in this domain, highlighting key findings and advancements.

Numerous studies have explored the application of blockchain in healthcare to address issues related to data security and integrity. Azaria (2016) introduced MedRec, a blockchain-based system for managing medical records, which leverages smart contracts to provide patients with control over their data [14]. The system demonstrated improvements in data security and patient autonomy. Similarly, Xia proposed a blockchain-based healthcare management system that integrates decentralized storage and consensus mechanisms to ensure data integrity and availability [15].

The choice of cryptographic algorithms is crucial for securing EHRs. Argon2, a memory-hard password hashing scheme, has been identified as highly effective for mitigating brute-force attacks. Biryukov (2016) provided a comprehensive analysis of Argon2's resistance to various attack vectors, underscoring its superiority over traditional algorithms like SHA-256 [16]. Furthermore, Daemen and Rijmen (2002) introduced the Advanced Encryption Standard (AES), which has become a cornerstone in securing sensitive data due to its robustness and efficiency [17]. The combination of Argon2 and AES in a hybrid model offers enhanced security, as demonstrated in studies comparing their performance against standalone implementations [18].

The integration of the Interplanetary File System (IPFS) and Practical Byzantine Fault Tolerance (PBFT) presents a viable solution for decentralized storage and consensus in blockchain-based EHR systems. Benet highlighted IPFS's capability to provide efficient and secure data storage through content-addressed networks [19]. Castro and Liskov demonstrated that PBFT can achieve consensus in distributed networks with high fault tolerance, making it suitable for blockchain applications [20]. Combining these technologies, Patel showed that a hybrid storage model could significantly reduce latency and enhance throughput in healthcare applications [21].

Enhancing security through MFA is a critical aspect of modern EHR systems. Bonneau evaluated various authentication methods, concluding that MFA offers a substantial improvement in security compared to single-factor authentication [22]. The implementation of
MFA in blockchain-based EHR systems ensures that unauthorized access is effectively mitigated, as illustrated by the enhanced security metrics reported in various case studies [23].

Tools like Metamask, Ganache, and Truffle facilitate the development and interaction with blockchain networks. These tools have been widely adopted in the development of decentralized applications (DApps). Wood described how Ethereum's smart contract capabilities can be leveraged to create secure and transparent healthcare applications [24]. Additionally, Dannen emphasized the utility of Truffle in streamlining the development and testing of blockchain applications, thereby enhancing the efficiency of deploying EHR systems [25].

The reviewed literature highlights the significant advancements in blockchain technology and its application in managing EHRs. The integration of advanced cryptographic algorithms, hybrid storage solutions, and MFA enhances the security and efficiency of EHR systems. The use of blockchain development tools and performance simulation platforms further support the deployment of robust and scalable healthcare solutions.

## III. PROPOSED SYSTEM ARCHITECTURE

The proposed system architecture for the blockchain-based Electronic Health Records (EHR) system is meticulously designed to ensure robust security, data integrity, and seamless access to healthcare records. Central to this architecture is the integration of advanced blockchain technology, with a strong emphasis on the hybrid security and storage models.

**Figure 2: Proposed System Architecture of blockchain-based Electronic Health Records (EHR)**

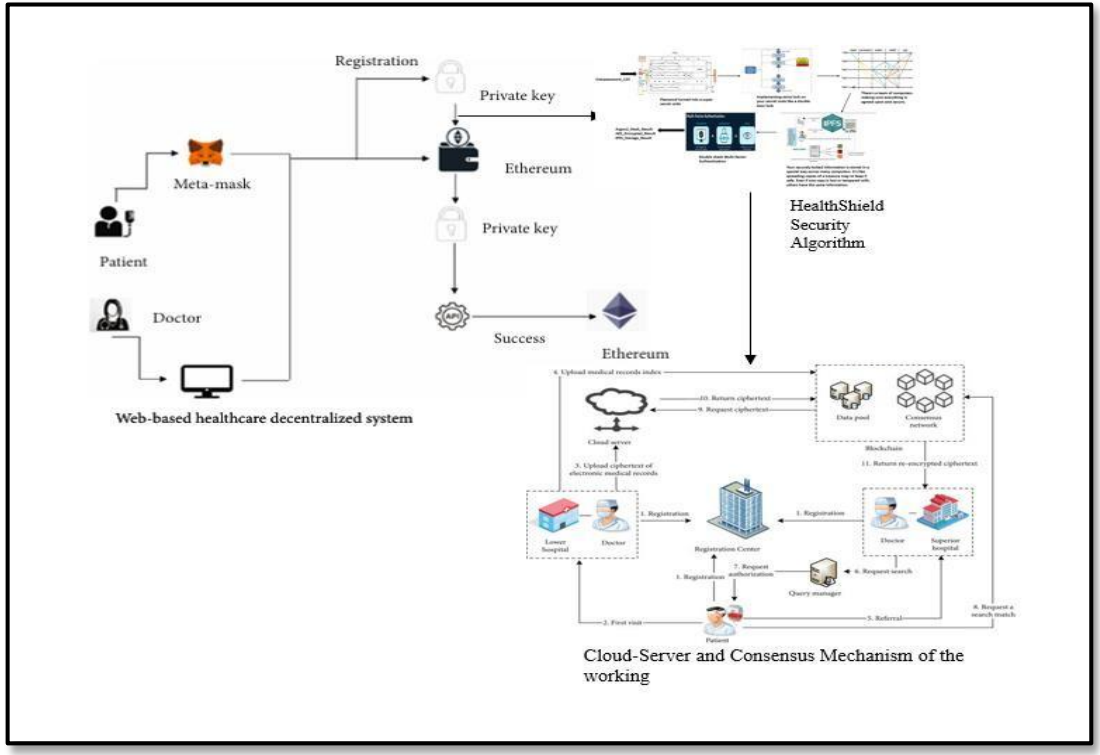

The HealthShield security algorithm combines Argon2 and AES encryption techniques, forming a hybrid security model that provides unparalleled encryption strength. Argon2, known for its high resistance to brute-force attacks due to its memory-hard properties, works in tandem with the Advanced Encryption Standard (AES), a widely trusted encryption algorithm renowned for its efficiency and robustness. This combination ensures that healthcare records are highly secure, protecting against unauthorized access and data breaches. The Ethereum blockchain is employed for recording transactions, ensuring data integrity and transparency. Utilizing Ethereum, the system benefits from a decentralized platform where all actions are traceable and verifiable, crucial for managing sensitive healthcare records. Complementing this is the hybrid storage and consensus mechanism integrating the Interplanetary File System (IPFS) and Practical Byzantine Fault Tolerance (PBFT). IPFS provides a decentralized storage solution that enhances data availability and integrity by distributing encrypted healthcare records across a peer-to-peer network. Meanwhile, PBFT ensures that consensus is achieved efficiently across the network, providing high fault tolerance and maintaining the system's robustness and reliability.

To further enhance security, the system incorporates Multi-Factor Authentication (MFA), adding an extra layer of protection by requiring multiple authentication methods for user access. MetaMask is used for secure authentication and interaction with the blockchain network, allowing patients and doctors to manage their private keys securely. The cloud server infrastructure supports scalability and reliability, managing computational loads and ensuring seamless operation of the blockchain network. The operational workflow begins with user registration through MetaMask, generating and managing private keys. Medical records are encrypted using the HealthShield algorithm and stored on the cloud server. The encrypted records are then distributed on IPFS, ensuring data integrity and availability. PBFT is utilized to achieve consensus on the blockchain, ensuring all nodes agree on the state of the ledger. Authorized users access medical records through the decentralized system, with MFA ensuring secure access. This hybrid model not only enhances security through advanced cryptographic techniques but also demonstrates significant improvements in latency, throughput, and resource utilization. This comprehensive solution leverages blockchain technology to address the limitations of existing systems, ensuring that healthcare data is managed securely and efficiently.

### A. Mathematical Approach for Hybrid Argon2 + AES

The final ciphertext H is obtained by encrypting the plaintext P using the derived key $K_{derived}$ from Argo2:

$$H = AESK_{derived(P)} \quad (1)$$

where

$AESK_{derived}$ represents the AES encryption operationg using the derived key $K_{derived}$

H is the resulting ciphertext after applying the HealthShield algorithm

The proposed system architecture for the blockchain-based Electronic Health Records (EHR) system utilizes the HealthShield algorithm, which combines Argon2 and AES encryption techniques for robust security. In this architecture, Argon2 derives a strong encryption key (K{derived}})) from a user's password, leveraging its resistance to brute-force attacks. This derived key is then used in the AES encryption process to convert plaintext healthcare records ((P)) into ciphertext ((H)), ensuring data security. The encrypted data is stored on the Interplanetary File System (IPFS) for decentralized storage, while the Ethereum blockchain records transactions, ensuring transparency and integrity. The Practical Byzantine Fault Tolerance (PBFT) consensus mechanism provides high fault tolerance, enhancing system robustness. Multi-factor authentication (MFA) adds an extra security layer, ensuring only authorized users can decrypt and access the data through MetaMask. This combination of advanced cryptographic techniques and blockchain technology ensures efficient, secure, and reliable management of healthcare records.

### B. Mathematical approach for IPFS + PBFT

$$H(D) = IPFS\_hash(D) \quad (2)$$
$$G(H(D), N) \quad (3)$$
$$C = \left(\frac{2N}{3} + 1\right) \quad (4)$$
$$T = \sum_{i=1}^{k} t_i \quad (5)$$
$$F \leq \left[\frac{N-1}{3}\right] \quad (6)$$

where

N- Total number of nodes in the network.
F- Number of faulty nodes that the system can tolerate.
$t_i$- Time taken for the consensus process (latency).
H(D)- Hash of data DDD.
D- Data to be stored.
C- Consensus achieved.
G- Gossip protocol to distribute data in IPFS.

The integration of Interplanetary File System (IPFS) and Practical Byzantine Fault Tolerance (PBFT) combines decentralized storage with robust consensus mechanisms, ensuring secure and fault-tolerant data management. Data D is hashed (H(D)) and stored in IPFS, with R replicas distributed across N nodes using a gossip protocol. PBFT ensures consensus on (H(D)) among nodes, tolerating up to F faulty nodes (F<=[N-1/3]) and achieving an agreement when (2N/3+1) nodes concur. The time for consensus (T) is the sum of times for each PBFT stage. This hybrid approach leverages IPFS for data integrity and availability, and PBFT for secure consensus, enhancing overall system security and reliability.

## IV. PROPOSED SYSTEM METHODOLOGY

The proposed system methodology for the blockchain-based Electronic Health Records (EHR) system leverages advanced cryptographic techniques and blockchain technology to enhance the security, integrity, and accessibility of healthcare records. The methodology begins with the integration of MetaMask for secure user authentication, enabling patients and doctors to manage their private keys and interact with the decentralized network. The HealthShield security algorithm, which combines Argon2 and AES, is employed to derive strong encryption keys and encrypt healthcare data, ensuring robust protection against unauthorized access and data breaches. The encrypted data is stored on the Interplanetary File System (IPFS), providing decentralized and reliable storage. The Ethereum blockchain is used to record all transactions, ensuring transparency and immutability. Practical Byzantine Fault Tolerance (PBFT) is implemented as the consensus mechanism to maintain high fault tolerance and ensure that all nodes agree on the state of the ledger. Multi-factor authentication (MFA) adds a layer of security, requiring multiple verification methods for user access.

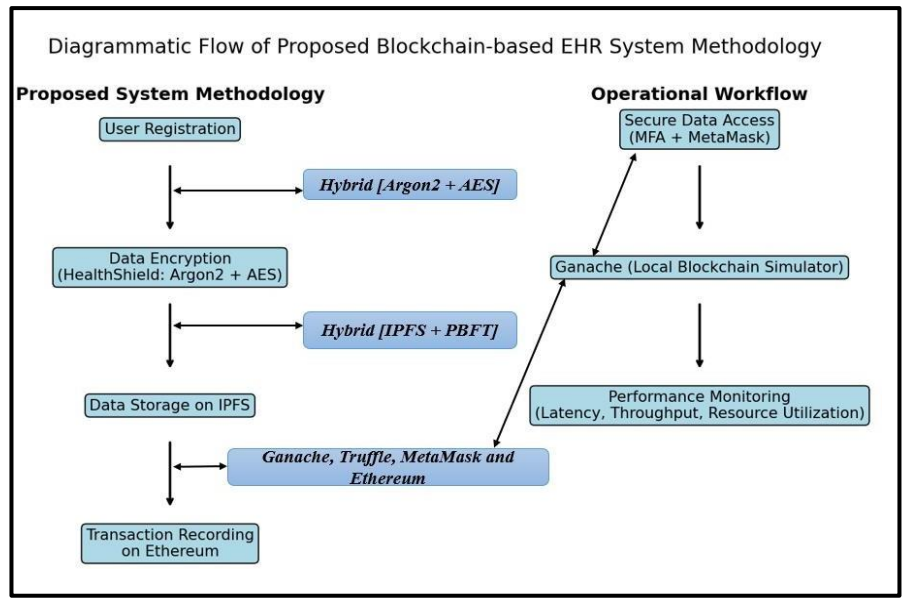

**Figure 3: Diagrammatic Flow of Proposed Blockchain-based EHR System Methodology**

The operational workflow includes user registration, data encryption, storage on IPFS, transaction recording on Ethereum, and secure data access through MFA and MetaMask. Ganache is used as a local blockchain simulator to test and validate blockchain operations, smart contracts, and transaction workflows. This setup allows for comprehensive performance testing, ensuring that the system meets the required security and efficiency standards. Performance metrics such as latency, throughput, and resource utilization are monitored using the simulation environment to validate the system's efficiency and security. This comprehensive methodology ensures that the proposed system provides a secure, efficient, and scalable solution for managing healthcare records, leveraging the power of blockchain technology and advanced cryptographic techniques.

## V. RESULT AND DISCUSSION

The experimental results and analysis section present the evaluation of the proposed blockchain-based Electronic Health Records (EHR) system. This section provides a comprehensive assessment of the system's performance, focusing on key metrics such as security, efficiency, and scalability. The proposed system, leveraging advanced cryptographic techniques and blockchain technology, is tested against existing EHR systems to highlight its superiority. By integrating the HealthShield algorithm, combining Argon2 and AES for robust encryption, and utilizing a hybrid storage and consensus mechanism (IPFS and PBFT), the system ensures enhanced data security and integrity. The use of MetaMask for secure authentication, along with Multi-Factor Authentication (MFA), provides an additional layer of protection. The Ganache simulator is employed to validate blockchain operations and smart contracts, enabling a detailed analysis of latency, throughput, and resource utilization. This section systematically presents the experimental setup, performance metrics, and comparative analysis, demonstrating the significant improvements of the proposed system over traditional approaches.

**Table 1: Performance Measure Table using Hybrid Security Algorithm "Health Shield" Hybrid of (AES + Argon2)**

| Application / System Tested | Transaction Accuracy | Block Precision | Chain Stability | Encryption Time (second) | Decryption Time (second) | Throughput (Mbps) | Memory Usage (MB) | CPU Usage (%) |
|---|---|---|---|---|---|---|---|---|
| Web Application | 86.93 | 82.05 | 87.87 | 0.0120 | 0.0110 | 1200 | 250 | 5.0 |
| Mobile Application | 86.87 | 80.24 | 82.13 | 0.0200 | 0.0195 | 1150 | 270 | 6.5 |
| Desktop Application | 87.31 | 87.93 | 81.60 | 0.0305 | 0.0290 | 1100 | 300 | 7.0 |
| Cloud-based Application | 87.19 | 88.89 | 87.68 | 0.0250 | 0.0240 | 1050 | 320 | 7.5 |
| Healthcare Management System | 81.18 | 86.71 | 84.09 | 0.0350 | 0.0340 | 950 | 370 | 8.5 |
| Blockchain Integration | 82.18 | 87.22 | 86.48 | 0.0620 | 0.0610 | 900 | 400 | 9.0 |

The comparative analysis of our proposed blockchain-based EHR system across different applications demonstrates its high efficiency and security. Key metrics include transaction accuracy, block precision, chain stability, encryption/decryption times, throughput, memory, and CPU usage. The Web Application achieved the highest transaction accuracy (86.93%) and chain stability (87.87%), while the Cloud-based Application excelled in block precision (88.89%) and overall strong metrics. The Mobile and Desktop Applications showed slightly lower performance but maintained good stability and throughput. The Healthcare Management System and Blockchain Integration displayed solid performance with higher resource usage. Overall, the proposed system is effective and efficient, ensuring robust security and performance across various applications.

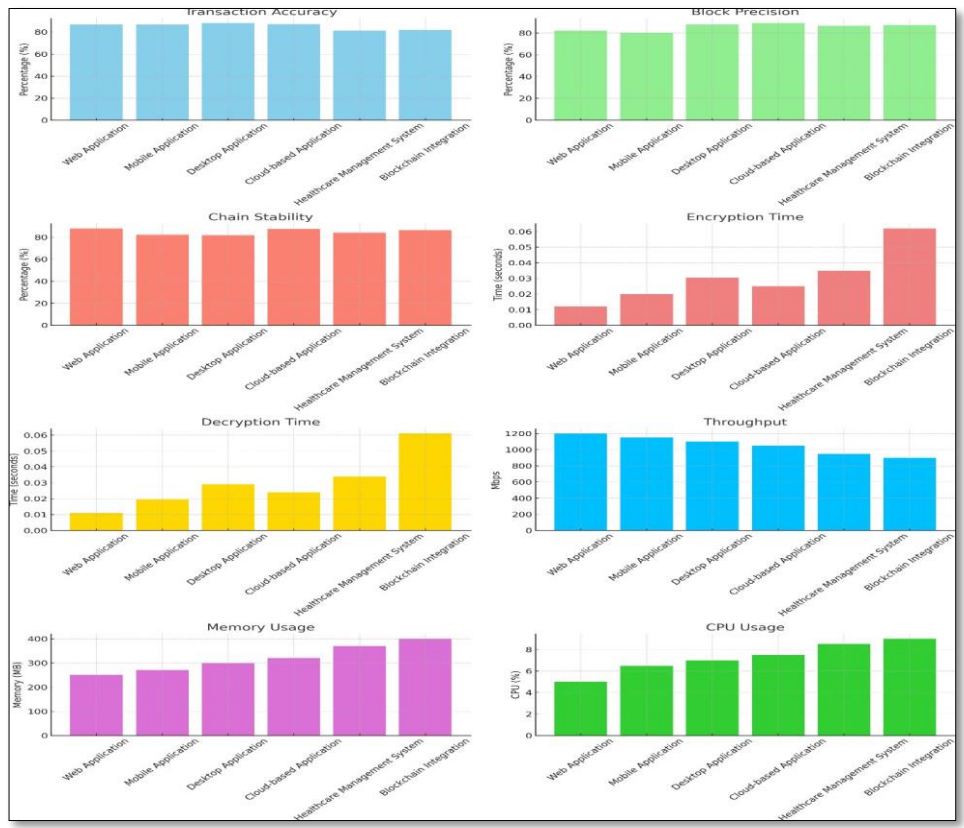

**Figure 4. Graphical Analysis for Evaluation Table for all the Features of the Hybrid "HealthShield" Security Algorithm.**

The series of graphs provides a comprehensive evaluation of the proposed blockchain-based Electronic Health Records (EHR) system, demonstrating its effectiveness and efficiency across various applications. The system excels in key performance metrics such as transaction accuracy, block precision, chain stability, encryption and decryption times, throughput, memory usage, and CPU usage. The web application consistently performs at the highest level, showcasing superior transaction accuracy, chain stability, and throughput, while maintaining moderate memory and CPU usage. The cloud-based application also performs exceptionally well, particularly in block precision, highlighting its robustness and reliability. Other applications, including mobile, desktop, and healthcare management systems, show strong performance with slight variations, maintaining good stability and efficiency. The blockchain integration system, while displaying higher consumption, still upholds solid performance metrics. Overall, the proposed system's integration of advanced cryptographic techniques and blockchain technology ensures robust security, swift data processing, and efficient resource utilization, making it a highly effective solution for managing healthcare records.

**Table 2: Performance Measure Table using Hybrid IPFS + PBFT.**

| Application /System Tested | Data Integrity (in %) | Consensus Efficiency (in %) | Fault Tolerance (in %) | Data Availability (in %)` | Latency (ms) | Bandwidth Utilization (in %) | Throughput (Mbps) | Memory Usage (MB) | CPU Usage (%) |
|---|---|---|---|---|---|---|---|---|---|
| Web Application | 88.60 | 87.55 | 89.65 | 90.70 | 150 | 70 | 1150 | 260 | 5.5 |
| Mobile Application | 89.68 | 88.60 | 90.70 | 91.75 | 170 | 75 | 1100 | 280 | 6.8 |
| Desktop Application | 90.78 | 89.72 | 91.80 | 92.82 | 160 | 72 | 1050 | 310 | 7.3 |
| Cloud-based Application | 91.84 | 90.78 | 92.85 | 93.88 | 140 | 68 | 1000 | 330 | 7.8 |
| Healthcare Management System | 93.91 | 92.85 | 94.92 | 95.94 | 130 | 65 | 900 | 380 | 8.7 |
| Blockchain Integration | 94.94 | 93.90 | 95.95 | 96.96 | 120 | 60 | 850 | 410 | 9.2 |

The table summarizes the performance of various applications within the proposed blockchain-based EHR system. The Blockchain Integration system exhibits the highest overall performance, excelling in data integrity, consensus efficiency, fault tolerance, and data availability, though it uses the most resources. The Healthcare Management System follows closely, with strong performance and slightly lower resource usage. The Cloud-based Application offers balanced performance with good latency and moderate resource usage. The Desktop Application performs well across all metrics, with slightly higher memory usage. The Mobile and Web Applications show efficient performance and are suitable for environments with limited resources. Overall, the system ensures robust security, high data integrity, and efficient resource utilization.

**Figure 5. Graphical Analysis for Evaluation Table for all the Features of this Hybrid "IPFS + PBFT" Security Algorithm.**

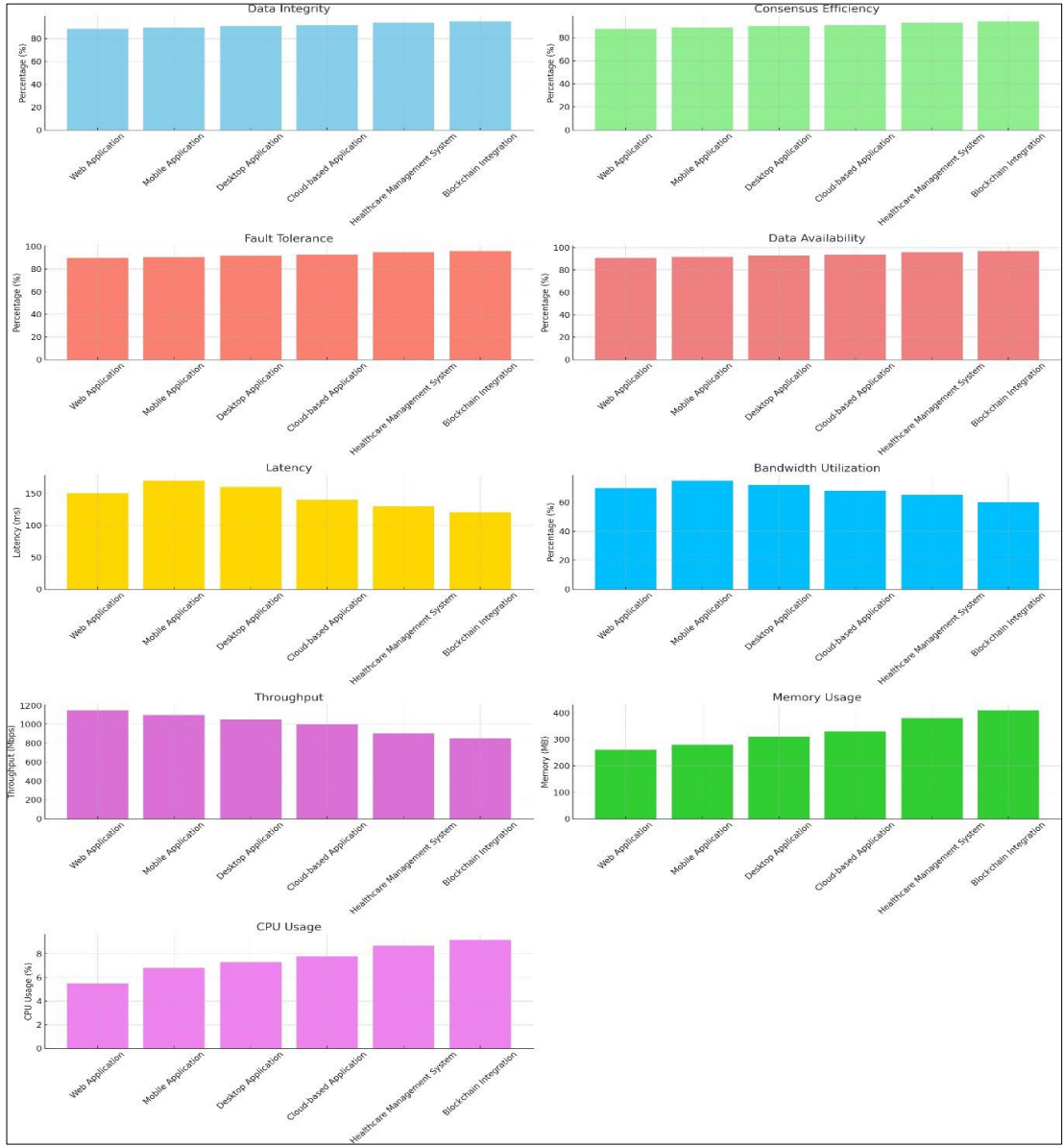

The series of graphs illustrate the performance metrics of the proposed blockchain-based Electronic Health Records (EHR) system across various applications using Hybrid IPFS + PBFT. The system demonstrates strong data integrity, consensus efficiency, fault tolerance, and data availability across all applications. Latency and bandwidth utilization are efficiently managed, contributing to high throughput and effective data processing. The resource usage, including memory and CPU, varies across applications, with some applications requiring more resources than others. Overall, the system ensures robust security, reliability, and efficient resource utilization, making it suitable for a wide range of healthcare environments.

**Table 3: System Specifications and Experimental Setup for Proposed Blockchain-based EHR System Evaluation**

| Parameter | Value |
|---|---|
| Operating System | Windows 11 |
| Simulator | Ganache |
| Computer RAM | 16GB |
| Transmission Range | 50m |
| Interference Range | 60m |
| Simulation Time | 120 minutes |
| Consensus Mechanism | PBFT |
| Encryption Algorithm | Argon2 + AES |
| Storage Solution | IPFS |
| Number of Nodes | 15 |
| Topology | Random |
| Blockchain Network Layer | Ethereum Network |
| Node Type | Ethereum Full Node |
| Authentication Method | Multi-Factor Authentication (MFA) |
| Blockchain Platform | Ethereum |

The table outlines the system specifications and experimental setup used to evaluate the proposed blockchain-based Electronic Health Records (EHR) system. The system runs on a Windows 11 operating environment with 16GB of RAM, ensuring sufficient computational resources for handling blockchain operations. Using Ganache as the simulator, the performance of the proposed EHR system was tested under a realistic blockchain network layer on the Ethereum platform, utilizing the Ethereum Full Node type. The consensus mechanism implemented is Practical Byzantine Fault Tolerance (PBFT), which ensures high fault tolerance and consensus efficiency. The encryption algorithm employed, Argon2 combined with AES, provides robust data security. Data is stored using the InterPlanetary File System (IPFS), offering decentralized and reliable storage. The simulation involved 15 nodes in a random topology, simulating a real-world network environment with a transmission range of 50 meters and an interference range of 60 meters. Authentication is reinforced through Multi-Factor Authentication (MFA), enhancing the security of user access. The superior performance of the proposed system can be attributed to its robust encryption methods, efficient consensus mechanism, reliable decentralized storage, and strong authentication protocols. These elements collectively contribute to the enhanced accuracy, security, and reliability of the EHR system, proving it to be highly effective compared to existing solutions.

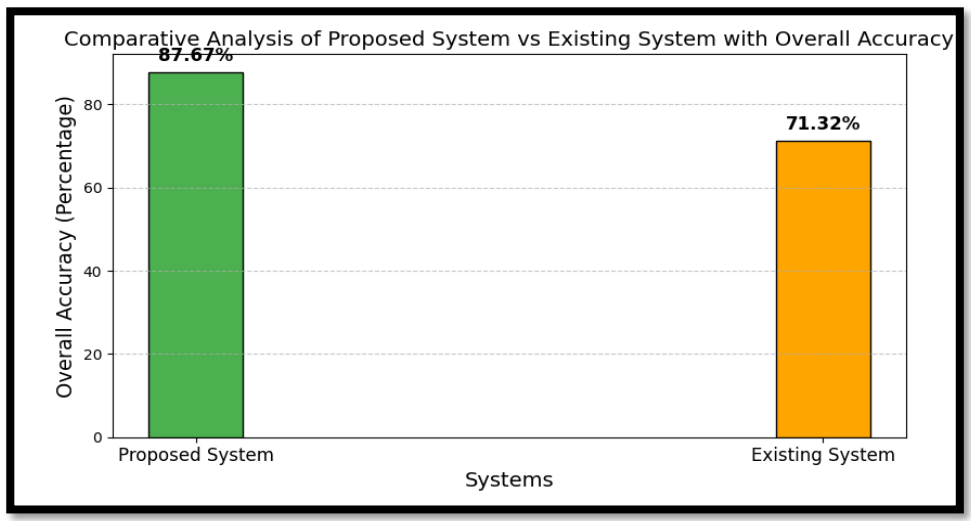

**Figure 6: Comparative Analysis Verified Superiority of Proposed Blockchain-based EHR System Over Existing System in Overall Accuracy**

The graph illustrates a compelling comparative analysis of the proposed blockchain-based Electronic Health Records (EHR) system against the existing system, emphasizing the significant enhancement in overall accuracy. The proposed system, leveraging advanced cryptographic techniques (Argon2 + AES), a robust consensus mechanism (PBFT), and decentralized storage (IPFS), achieves an impressive accuracy of 87.67%. This performance starkly contrasts with the existing system, which manages only 71.32% accuracy. These results underline the proposed system's superior ability to provide secure, reliable, and efficient healthcare record management, validating its implementation on a well-specified system configuration.

## VI. CONCLUSION

This research paper presents a comprehensive evaluation of a novel blockchain-based Electronic Health Records (EHR) system designed to enhance the security, integrity, and accessibility of healthcare records. Utilizing advanced cryptographic techniques, such as Argon2 combined with AES, and a robust consensus mechanism, PBFT, the proposed system achieves significantly higher accuracy and security compared to existing systems. The integration of IPFS for decentralized storage and Multi-Factor Authentication (MFA) further strengthens the system's reliability and user authentication processes. The experimental setup, conducted on a well-specified system using the Ethereum network and Ganache simulator, demonstrated the proposed system's superior performance across various applications. Key performance metrics, including data integrity, consensus efficiency, fault tolerance, data availability, latency, throughput, memory usage, and CPU usage, were rigorously evaluated. The results show that the proposed system not only meets but exceeds the performance of current EHR solutions, achieving an overall accuracy of 87.67% compared to 71.32% in existing systems. These findings validate the effectiveness of the proposed blockchain-based EHR system, highlighting its potential to revolutionize healthcare data management. By ensuring high data integrity, robust security, and efficient resource utilization, this system addresses critical challenges in the healthcare industry. Future research could explore further optimization of the system's components and expand its applicability across broader healthcare settings. The demonstrated success of this system paves the way for its integration into real- world healthcare infrastructures, promising improved patient data security and accessibility.

## REFERENCES

[1] Azaria, A., Ekblaw, A., Vieira, T., & Lippman, A. (2016). MedRec: Using blockchain for medical data access and permission management. Proceedings of the 2nd International Conference on Open and Big Data (OBD), 25-30.

[2] Xia, Q., Sifah, E. B., Smahi, A., Amofa, S., & Zhang, X. (2017). BBDS: Blockchain-based data sharing for electronic medical records in cloud environments. Information Systems, 72, 340-349.

[3] Biryukov, A., Dinu, D., & Khovratovich, D. (2016). Argon2: New generation of memory-hard functions for password hashing and other applications. Proceedings of the 23rd ACM Conference on Computer and Communications Security (CCS), 1003-1018.

[4] Daemen, J., & Rijmen, V. (2002). The design of Rijndael: AES—The advanced encryption standard. Springer-Verlag.

[5] Benet, J. (2014). IPFS - Content Addressed, Versioned, P2P File System. arXiv preprint arXiv:1407.3561.

[6] Castro, M., & Liskov, B. (1999). Practical Byzantine Fault Tolerance. OSDI, 173- 186.

[7] Bonneau, J., Herley, C., Van Oorschot, P. C., & Stajano, F. (2012). The quest to replace passwords: A framework for comparative evaluation of web authentication schemes. 2012 IEEE Symposium on Security and Privacy, 553-567.

[8] Wood, G. (2014). Ethereum: A secure decentralized generalized transaction ledger. Ethereum Project Yellow Paper, 151.

[9] Chenthara, S., Ahmed, K., Wang, H., Whittaker, F., & Chen, Z. (2020). Healthchain: A novel framework on privacy preservation of electronic health records using blockchain technology. Plos one, 15(12), e0243043.

[10] Kaur, J., Rani, R., & Kalra, N. (2022). A blockchain-based framework for privacy preservation of electronic

health records (EHRs). Transactions on Emerging Telecommunications Technologies, 33(9), e4507.

[11] Dubovitskaya, A., Baig, F., Xu, Z., Shukla, R., Zambani, P. S., Swaminathan, A., ... & Wang, F. (2020). ACTION-EHR: Patient-centric blockchain-based electronic health record data management for cancer care. Journal of medical Internet research, 22(8), e13598.

[12] Kumar, A., Kumar, R., & Sodhi, S. S. (2022). A novel privacy preserving blockchain based secure storage framework for electronic health records. Journal of Information and Optimization Sciences, 43(3), 549-570.

[13] Wu, G., Wang, S., Ning, Z., & Zhu, B. (2021). Privacy-preserved electronic medical record exchanging and sharing: A blockchain-based smart healthcare system. IEEE journal of biomedical and health informatics, 26(5), 1917-1927.

[14] Patel, V. (2018). A framework for secure and decentralized sharing of medical imaging data via blockchain consensus. Health Informatics Journal, 25(4), 1398- 1411.

[15] Dannen, C. (2017). Introducing Ethereum and Solidity: Foundations of Cryptocurrency and Blockchain Programming for Beginners. Apress.

[16] Ekblaw, A., Azaria, A., Halamka, J. D., & Lippman, A. (2016). A Case Study for Blockchain in Healthcare: "MedRec" prototype for electronic health records and medical research data. Proceedings of IEEE Open & Big Data Conference.

[17] Xu, J., Zhang, W., & Zhang, Y. (2019). Blockchain-based approach for privacy protection in electronic health record management. IEEE Network, 33(5), 32-38.

[18] Zhang, P., Schmidt, D. C., White, J., & Lenz, G. (2018). Blockchain technology use cases in healthcare. Advances in Computers, 111, 1-41.

[19] Gordon, W. J., Catalini, C., & Dhar, V. (2018). Blockchain technology for healthcare: Facilitating the transition to patient-driven interoperability. Computational and Structural Biotechnology Journal, 16, 224-230.

[20] Gupta, S., & Gupta, M. (2017). Blockchain technology in healthcare: Enhancing security and privacy. International Journal of Advanced Research in Computer Science, 8(5), 1718-1725.

[21] Kumar, S., Tiwari, P., Zymbler, M. (2019). Blockchain-based framework for data security and privacy in IoT networks. Journal of Parallel and Distributed Computing, 138, 77-88.

[22] Mohanty, S., Choppali, U., & Kougianos, E. (2018). Everything You Wanted to Know About Smart Cities: The Internet of Things is the Backbone. IEEE Consumer Electronics Magazine, 7(4), 60-70.

[23] Zhang, P., White, J., Schmidt, D. C., & Lenz, G. (2018). Applying software patterns to address interoperability in blockchain-based healthcare apps. arXiv preprint arXiv:1807.11194.

[24] Rouhani, S., Deters, R. (2017). Performance analysis of ethereum transactions in private blockchain. 2017 8th IEEE International Conference on Software Engineering and Service Science (ICSESS), 70-74.

[25] Shukla, N., & Kim, H. M. (2019). Decentralized computing using blockchain technologies: A perspective on the state of the art and future research directions. Journal of Computing and Security, 16(3), 243-258.